\documentclass[conference,compsoc]{IEEEtran}
\ifCLASSOPTIONcompsoc
  \usepackage[nocompress]{cite}
\else
  \usepackage{cite}
\fi
\usepackage{xcolor}

\usepackage{algorithm}
\usepackage{algpseudocode}

\usepackage{amssymb,amsmath}

\ifCLASSINFOpdf
\else
\fi
\usepackage{epsfig}

\usepackage{url}

\usepackage{setspace}
\usepackage{lipsum}

\begin{document}
%
\title{{\huge $\delta$}-MAPS: From spatio-temporal data to a weighted and lagged
network between functional domains}



%
\author{\IEEEauthorblockN{Ilias Fountalis\IEEEauthorrefmark{1},
Annalisa Bracco\IEEEauthorrefmark{2},
Bistra Dilkina\IEEEauthorrefmark{3}, 
Constantine Dovrolis\IEEEauthorrefmark{1} and
Shella Keilholz\IEEEauthorrefmark{4}}
\IEEEauthorblockA{\IEEEauthorrefmark{1}School of Computer Science, Georgia Tech}
\IEEEauthorblockA{\IEEEauthorrefmark{2}School of Earth and Atmospheric Sciences, Georgia Tech\\}
\IEEEauthorblockA{\IEEEauthorrefmark{3}School of Computational Science and Engr., Georgia Tech}
\IEEEauthorblockA{\IEEEauthorrefmark{4}Dept. of Biomedical Engr., Georgia Tech and Emory\\ Email: fountalis@gatech.edu}}


\maketitle

\begin{abstract}
We propose $\delta$-MAPS, a method that analyzes spatio-temporal data 
first to identify the distinct spatial components of the underlying system,
referred to as ``domains'', and second to infer the connections between them. 
A domain is a spatially contiguous region of highly correlated temporal activity. 
The core of a domain is  a point or subregion at which a metric of local homogeneity is 
maximum across the entire domain. 
We compute a domain as the maximum-sized set of spatially contiguous cells that include the detected core
and satisfy a homogeneity constraint, expressed in terms of the average pairwise cross-correlation 
across all cells in the domain.  Domains may be spatially overlapping.
Different domains may have correlated activity, potentially at a lag, because of direct or indirect interactions.
The proposed edge inference method examines the statistical significance of each lagged 
cross-correlation between two domains, infers a range of lag values for each edge,
and assigns a weight to each edge based on the covariance of the two domains. 
We illustrate the application of $\delta$-MAPS on data from two domains: climate science 
and neuroscience. 

\end{abstract}


%
\IEEEpeerreviewmaketitle

\section{Introduction} \label{sec:introduction}
Spatio-temporal data become increasingly prevalent and important for both science (e.g., climate,
systems neuroscience, seismology) and enterprises (e.g., the analysis of geotagged social media activity).
The spatial scale of the available data is often determined by an arbitrary grid, 
which is typically larger than the true dimensionality of the underlying system. 
One major task is to identify the distinct semi-autonomous components of this system 
and to infer their (potentially lagged and weighted) interconnections from the available spatio-temporal data. 
Traditional dimensionality reduction methods, such as PCA, ICA or clustering, have been successfully
used for many years but they have known limitations when the objective is to infer the 
functional network between all spatial components of the system.

We propose $\delta$-MAPS, an inference method that first identifies these spatial components, 
referred to as ``domains'', and then the connections between them (\S\ref{sec:deltamaps}). 
Informally, a {\em functional domain} (or simply {\em domain}) is a spatially
contiguous region that somehow participates in the same dynamic effect or function.
The exact mechanism that creates this effect or function varies across application domains;
however, the key idea is that 
{\em the functional relation between the grid cells of domain results in highly correlated temporal activity}. 
If we accept this premise, it follows that we should be able to identify the ``epicenter''
or {\em core of a domain} as a point (or subregion) at which the local homogeneity is 
maximum across the entire domain. 
Instead of searching for the discrete boundary of a domain, which may not exist in reality, we compute 
a domain as the {\em maximum possible set} of spatially contiguous cells that include the detected core, 
and that satisfy a homogeneity constraint, expressed in terms of the 
average pairwise cross-correlation across all cells in the domain.
Domains may be spatially overlapping.
Also, some cells may not belong to any domain.

After we identify all domains,  $\delta$-MAPS infers a functional network between them. 
Different domains may have correlated activity, potentially at a lag, 
because of direct or indirect interactions.
The proposed edge inference method examines the statistical significance of each lagged 
cross-correlation between two domains, applies a multiple-testing process to control 
the rate of false positives, infers a range of potential lag values for each edge,
and assigns a weight to each edge based on the covariance of the corresponding two domains. 

$\delta$-MAPS is related to clustering, parcellation (or regionalization), network 
community detection, multivariate statistical methods for dimensionality reduction
such as PCA and ICA, as well as functional network and lag inference methods.  
However, as we discuss in \S\ref{sec:relatedwork} and show with synthetic data experiments 
in \S\ref{sec:validation}, $\delta$-MAPS is also significantly different than all these methods.  
$\delta$-MAPS does not require the number of domains as an input parameter,   
the resulting domains are spatially contiguous and potentially overlapping, 
and the inferred connections between domains can be lagged and positively or negatively weighted.
Further, the distinction between grid cells that are correlated within the same domain 
and grid cells that are correlated across two distinct domains allows 
$\delta$-MAPS to separate between local diffusion (or dispersion) phenomena and remote interactions 
that may be due to underlying structural connections (e.g., a white-matter fiber between two brain regions).

We illustrate the application of $\delta$-MAPS on data from two domains: climate science 
(\S\ref{sec:appclimate}) and neuroscience (\S\ref{sec:appfmri}). 
First, the sea-surface temperature (SST) climate network identifies some well-known climate 
``tele-connections'' (such as the lagged connection between the El Ni$\tilde{n}$o Southern Oscillation
and the Indian ocean) but it also captures less well-known lagged connections that deserve further investigation by the domain experts.  
Second, the analysis of resting-state fMRI cortical data confirms the presence of three well-known functional
brain ``networks'' (default-mode, occipital, and motor/somatosensory), and shows that the 
cortical network includes a {\em backbone} of relatively few regions that are densely interconnected.

\section{Related Work} \label{sec:relatedwork}

A common approach to reduce the dimensionality of spatio-temporal data is to apply PCA (standard or 
rotated) or ICA techniques.   
For instance, in climate science, PCA (also known as Empirical Orthogonal Function (EOF) analysis) has been 
used to identify teleconnections between distinct climate regions \cite{von2001statistical}.
The orthogonality between PCA components complicates the interpretation of the results making it 
difficult to identify the distinct underlying modes of variability and to separate their effects, 
as clearly discussed in \cite{dommenget2002cautionary}.
ICA analysis is more common in the neuroscience literature, aiming to identify independent rather than orthogonal
components \cite{hyvarinen1999fast}. 
However, ICA does not provide a relative significance for each component, and the number of independent 
components should be chosen based on some additional information about the underlying system.

Another broad family of spatio-temporal dimensionality reduction methods is based on unsupervised clustering. 
Such algorithms can be  grouped into region-growing (e.g., \cite{blumensath2012resting,lu2003region}),
spectral (e.g., the NCUT method often applied in fMRI analysis \cite{craddock2012whole,van2008normalized} -- but also 
see a discussion of their limitations \cite{baldassano2015parcellating}),
hierarchical (e.g., \cite{blumensath2013spatially,thirion2014fmri}), 
probabilistic (e.g., \cite{baldassano2015parcellating}) or density based methods \cite{kawale2013graph}. 
These groups of algorithms are quite different but they share some common characteristics: 
the resulting clusters may not be spatially contiguous \cite{steinbach2003discovery,van2008normalized}, 
every grid cell needs to belong to a cluster (potentially excluding only outliers) 
\cite{blumensath2012resting,lu2003region}, 
and the number of clusters is often required as an input parameter 
\cite{craddock2012whole,blumensath2013spatially} - none of these algorithms account for the fact that 
clusters may overlap. In particular, 
the lack of spatial contiguity makes it hard to distinguish between correlations due 
to spatial diffusion (or dispersion) phenomena from correlations that are due to remote 
(structural) interactions between distinct effects.

An approach of increasing popularity is to 
first construct a correlation-based network between individual grid cells,
after pruning cross-correlations that are not statistically significant -- see \cite{kramer2009network}.
Then, some of these methods analyze the (binary or weighted) cell-level network directly based on 
various centrality metrics, k-core decomposition, spectral analysis, etc. (e.g., \cite{donges2009backbone,van2011rich}) 
or they first apply a community detection algorithm (potentially able to detect overlapping communities, e.g.,  
\cite{ahn2010link,lancichinetti2011finding,palla2005uncovering}) 
on the cell-level network and then analyze the resulting communities in terms of size, density, location, overlap, etc. 
(e.g., \cite{mcguire2014community,power2011functional,steinhaeuser2010exploration,steinhaeuser2011complex}). 
A community however may group together two regions that are, first, not spatially contiguous, and second, 
different in terms of how they are connected to other regions; an instance of this issue is illustrated in  
Fig.~\ref{figEOFSlinksComms}-C in the context of climate data analysis.

\section{{\large $\delta$}-MAPS} \label{sec:deltamaps}


The input data is generated from a {\em spatial field} $\mathbf{X}(t)$ sampled on an arbitrary {\em grid} $G$.
This grid can be modeled as a planar graph $G(V,E)$, where each vertex in $V$ is a grid cell
and each edge in $E$ represents the spatial adjacency between two neighboring cells. 
A set of cells $A \subseteq V$ is {\em spatially contiguous}, denoted by $I_G(A)$=1, 
if it forms a connected component in $G$. 

The {\em $K$-neighborhood} of a cell $i$, denoted by $\Gamma_K(i)$, includes $i$ 
and the set of $K$ nearest neighbors to $i$
according to an appropriate spatial distance metric (e.g., geodesic distance for climate data, Euclidean distance for fMRI data).
The $K$-neighborhood of a cell is always spatially contiguous. 

Each grid cell $i$ is associated with a time series $x_i(t)$ of length $T$ ($t \in \{1,\dots T\}$). 
We assume that $x_i(t)$ is sampled from a stationary signal and denote by $\tilde{\mu}_i$ and 
$\tilde{\sigma}_i^2$ its sample mean and variance, respectively.  
The similarity between the activity of two cells $i$ and $j$ is measured 
with Pearson's cross-correlation at zero-lag,
\begin{equation} \label{EqPearson}
r_{i,j} = \frac{\sum_{t=1}^{T} (x_i(t)-\tilde{\mu}_i)(x_j(t)-\tilde{\mu}_j) } {T \, \tilde{\sigma}_i\tilde{\sigma}_j} \, .
\end{equation} 
Other similarity metrics could be used instead. 

The {\em local homogeneity at cell $i$} is defined as the average pairwise cross-correlation between 
the $K+1$ cells in $\Gamma_K(i)$,  
\begin{equation} \label{local-corr}
\hat{r}_{K}(i) = \frac{\sum_{m \neq n\in \Gamma_{K}(i)} r_{m,n}} {K \, (K+1)} \, .
\end{equation} 
Similarly, we define the {\em homogeneity of a set of cells $A$} as the average pairwise 
cross-correlation between all distinct cells in $A$,   
\begin{equation} \label{avg-corr}
\hat{r}(A) = \frac{\sum_{m \neq n\in A} r_{m,n}} {|A|\, (|A|-1)} \, .
\end{equation} 

\subsection{Functional domains}
Intuitively, a {\em domain} $A$ is a spatially 
contiguous set of cells that somehow participate in the same dynamic effect or function.
The exact mechanism that creates this effect or function varies across application domains;
however, the key premise is that {\em the functional relation between the cells of domain $A$ results
in highly correlated temporal activity (at zero-lag), and thus high values of the homogeneity metric $\hat{r}(A)$.}
A given {\em homogeneity threshold $\delta$} examines if the homogeneity 
of $A$ is sufficiently high, i.e., a domain $A$ must have $\hat{r}(A)>\delta$.
(the selection of $\delta$ is discussed later in this section).

If we accept this premise, it follows that we should be able to identify the ``epicenter'' 
or {\em core of a domain A} as a cell $i \in A$ at which the local homogeneity $\hat{r}_{K}(i)$ is 
maximum across all cells in $A$ (and certainly larger than $\delta$).  
In general, the core of a domain may not be a unique cell. 

More formally now, suppose that we know that cell $c$ is in the core of a domain. 
The {\em domain $A$ rooted at $c$} has to satisfy the following three properties: 
it should include cell $c$, be spatially contiguous, and have higher homogeneity than $\delta$:
\begin{equation} \label{delta-constraint}
c \in A, \quad I_G(A)=1, \quad \hat{r}(A) > \delta  \, .
\end{equation}

A domain may not have sharp spatial boundaries; instead, it 
may gradually ``fade'' into other domains or regions dominated by noise. 
So, instead of searching for the discrete boundary of a domain,
it is more reasonable to compute a domain as the {\em largest possible set of 
cells} that satisfies the previous three constraints. 

\noindent {\bf Domain identification problem: Given the field $\mathbf{X}(t)$ on the spatial grid 
$G$, a core cell $c$, and the threshold $\delta$, the domain $A(c)$ is a maximum-sized set 
of cells that satisfies the three constraints of (\ref{delta-constraint}).}  
In Appendix-1 we prove that the decision version of this problem is NP-Hard.

A given spatial field $\mathbf{X}(t)$  may include several domains. The number of identified domains,
denoted by $N$, depends on the threshold $\delta$. 
Domains may be spatially overlapping; this is the case when the cells of a region are
significantly correlated with two or more distinct domain cores. 
Also, some cells of the grid may not belong
to any domain, meaning that their signal can be thought of as mostly noise (at least for the given 
value of $\delta$).  
Decreasing $\delta$ will typically result in a larger number of detected domain cores.
Further, as $\delta$ decreases, the spatial extent of each domain will typically increase,
resulting in larger overlaps between nearby domains. 

$\delta$ can simply be a user-specified parameter for the minimum required average cross-correlation 
within a domain. 
Another way is to calculate $\delta$ based on a statistical test for the 
significance of the observed zero-lag cross-correlations. 
A summary of this method is given next (described in more detail in Appendix-2). 
We start with a random sample of pairs of grid cells. 
We then apply the statistical test described in \S\ref{sec:network-infer} (see Equations~
\ref{EqBarlettVariance} and \ref{test-corr}) to examine if the zero-lag cross-correlation
between each of these pairs passes a given significance level $\alpha$ 
(set to $10^{-2}$ unless specified otherwise).  
$\delta$ is then set to the average of the statistically significant cross-correlations in that sample. 
The rationale is that the average pairwise cross-correlation among cells that belong to the 
same domain should be higher than a sample average of statistically significant cross-correlations 
between cells that can be anywhere on the grid.

\subsubsection{Algorithm for domain identification}
Given the NP-Hardness of the previous problem, we propose a greedy algorithm that runs in two phases. 
In the first phase, we identify a set of cells, referred to as {\em seeds}; each seed is a candidate
core for a domain. 
In the second phase, each seed is initially considered as a distinct domain. 
Then, an iterative and greedy algorithm attempts to identify the largest possible domains
that satisfy the three constraints of (\ref{delta-constraint})
through a sequence of {\em expansion} and {\em merging} operations.
The two phases are described next, while
the complete pseudocode is presented in Appendix-3. The source code (including supporting
documentation) will be available on GitHub before the final publication of this paper.
\\

{\large {\em Seed selection.}} Recall that the core of a domain is a cell of maximum 
local homogeneity across all cells of that domain. 
So, one way to detect {\em potential} core cells, while the domains are still unknown, 
is to identify points at which the homogeneity field $\hat{r}_K(i)$ is locally maximum.
Specifically, cell $i$ is a seed if $\hat{r}_K(i)> \delta$ and $\hat{r}_K(i)\geq \hat{r}_K(j)$
$\forall j \in \Gamma_K(i)$. 
Let $S$ be the set of all identified seeds. 

In general, a single domain may produce more than one seed because the local homogeneity field can 
be noisy and so it may include multiple local maxima, greater than $\delta$.
Further, additional seeds can appear in regions where domains overlap.
Consequently, it is necessary to include a merging operation in which two or more seeds
are eventually merged into the same domain.  

Note that as $K$ decreases, the local homogeneity field becomes more noisy and so we may detect
more seeds in the same domain.
On the other hand, larger values of the neighborhood size $K$ can oversmooth the homogeneity field, 
removing seeds and potentially hiding entire domains.  
The latter is more likely if the spatial extent of a domain is smaller than $K$+1 cells.
This observation implies that the spatial resolution of the given grid sets a lower bound 
on the size of the functional domains that can be detected. 
\\

{\large{\em Domain-merging operation.}} 
Two candidate domains $A$ and $B$ can be merged if they are spatially contiguous and 
if the homogeneity of their union is sufficiently high, i.e., $\hat{r}(A\cup B) > \delta$.  
Whenever there is more than one pair of domains that can be merged, we greedily choose
the pair with the maximum union homogeneity; this greedy choice makes the merged
domain more likely to expand further. 

The merging operation is performed initially on the set of seeds $S$.
It is also performed after each domain-expansion operation, whenever it is possible to do so.  
\\

{\large {\em Domain-expansion operation.}}
A domain $A$ is expanded by considering all cells that are adjacent to $A$,
and selecting the cell $i$ that maximizes $\hat{r}(A \cup \{i\})$; again, this greedy choice
makes the expanded domain more likely to expand further. 

The expansion operation is repeated in rounds. 
At the start of each round, domains are sorted in decreasing order of homogeneity.
Then, each domain is expanded by one cell at a time, as previously described, in that order.  
After every expansion operation, we check whether one or more merging operations are possible.
A round is complete when we have attempted to expand each domain once. 

A domain can no longer expand if that would violate the homogeneity constraint $\delta$ or if there
are no other adjacent cells that can be added into the domain.  
The domain identification algorithm terminates when no further expansion or merging operations are possible.

\subsection{The domain network} \label{sec:network-infer}
Given the $N$ identified domains $V_\delta = \{A_1, \dots A_N\}$, the next step is to construct a
network $G_\delta(V_{\delta},E_{\delta})$ between domains.   
Different domains may have correlated activity because of direct or indirect interactions.  
We refer to $G_\delta$ as  a {\em functional network} 
to emphasize that the edges between domains are based on functional activity and correlations
instead of structural or physical connections (``structural network'') 
or causal interactions (``effective network'').  

We associate a {\em domain-level signal} $X_A(t)$ with each domain $A$. 
The definition of this signal depends on the specific application field.  
For instance, when we analyze climate anomaly time series, the domain-level signal is defined
as the {\em cumulative anomaly} across all cells of that domain, where the contribution of each signal
is weighted by the relative size of that cell (it depends on the cell's latitude).
For fMRI data, the domain-level signal is defined as the {\em average BOLD signal} across the cells
of that domain. 

\begin{figure}[htpb]
\centering
\epsfig{file=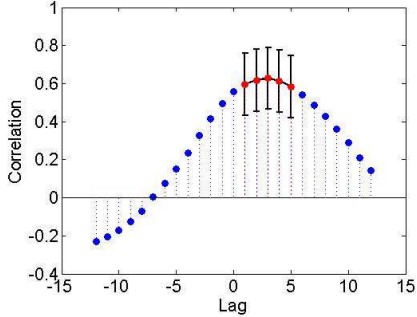,scale = 0.30}
\caption{Correlogram between two climate time series for a lag range of $\pm 12$ months. 
We show the significant correlations for a false discovery rate $q = 10^{-3}$ with red. 
The error bars correspond to $\pm$ one standard deviation, as estimated by Eq.~(\ref{EqBarlettVariance}). }
\label{figNetInfCorr}
\end{figure}

Two different domains may be located at some distance, and so they may be correlated at a non-zero lag $\tau$.  
For this reason, we examine if there is a significant cross-correlation between different domains
over a range of lags ($-\tau_{max} \leq \tau \leq \tau_{max}$).
The sample cross-correlation between domains $A$ and $B$ at a lag $\tau$ can be estimated as: 
\begin{equation} \label{PearsonEqLag}
r_{A,B}(\tau) = \frac{ \sum_{t = 1}^{T-\tau} (X_A(t) - \tilde{\mu}_A)(X_B(t+\tau) - \tilde{\mu}_B) } {T \tilde{\sigma}_A \tilde{\sigma}_B } \, ,
\end{equation}
where $\tilde{\mu}_A$ and $\tilde{\sigma}_A$ denote sample mean and standard deviation estimates, respectively.
The selection of $\tau_{max}$ should be large enough to include the typical signal propagation delays
in the underlying system but at the same time it should be much lower than $T$.
The $2\tau_{max}+1$ cross-correlations for a pair of domains can be represented with a {\em correlogram}; an example 
based on climate sea-surface temperature data (see \S\ref{sec:appclimate}) is shown in Fig.~\ref{figNetInfCorr}.

The next step is to examine the statistical significance of the measured cross-correlation between two domains 
$A$ and $B$.
Two uncorrelated signals can still produce a considerable sample cross-correlation if they have a strong 
auto-correlation structure.  
This is captured by the Bartlett's formula \cite{box2011time}, which is an estimator for the variance
of $r_{A,B}(\tau)$ (for a fixed value of $\tau$).
Under the null-hypothesis that the domain-level signals of $A$ and $B$ are uncorrelated, 
\begin{equation}\label{EqBarlettVariance}
\mathrm{Var}[r_{A,B}(\tau)] = \frac{1}{T-\tau} \sum_{\tau_k = -T}^{T}  r_{A,A}(\tau_k) \, r_{B,B}(\tau_k) \, ,
\end{equation}
where $r_{A,A}(\tau_k)$ is the autocorrelation of the time series of domain $A$ at lag $\tau_k$.

Under the previous null-hypothesis, the expected value of $r_{A,B}(\tau)$ is zero
and the following statistic approximately follows the standard normal distribution $N(0,1)$:
\begin{equation} \label{test-corr}
z_{A,B}(\tau) = \frac{ r_{A,B}(\tau)}{\sqrt{\mathrm{Var}[r_{A,B}(\tau)]}} \, .
\end{equation}
The approximation is due to the fact that $r_{A,B}(\tau)$ is bounded between $[-1,1]$.
So, we can now perform hypothesis testing for every pair of domains, computing a corresponding $p$-value based on $z$.

Given that there may be several domains in $G_{\delta}$, we need to control the number of false
positive edges that may result from the multiple testing problem. We do so using the False Discovery Rate (FDR)
method of Benjamini and Hochberg \cite{benjamini1995controlling}. 
Specifically, given $N$ domains, we need to perform $M=\frac{N(N-1)}{2}\,(2\tau_{max}+1)$ tests (for each potential
edge and for each possible lag value), and  compute the $p$-value for each test, based on (\ref{test-corr}). 
Given a False Discovery Rate $q$ (the expected value of the fraction of tests that are false positives), 
the Benjamini-Hochberg procedure ranks the $M$ p-values ($p_i$ becomes the $i$'th lowest $p$-value) 
and only keeps the first $m<M$ tests (edges), where $p_m$ is the highest $p$-value such that $p_m < q \, m / M$. \footnote{ This formula assumes that the p-values are independent (which is often not true in practice). The case of correlated p-values can be handled replacing $q$ qith $q/\sum_{i=1}^m 1/i$, but that approach is very conservative, resulting in many false negatives \cite{reiner}.}
\\

{\large {\em Lag inference and edge directionality.}}
We infer the domain-level network $G_{\delta}$ as follows. Two domains $A,B \in V_\delta$ are connected
if there is at least one lag value at which the cross-correlation $r_{A,B}(\tau)$ has passed the FDR test. 
The standard approach in {\em lag inference} is to consider the lag value $\tau^*$ that maximizes the absolute
cross-correlation,
\begin{equation} \label{max-lag}
\tau^*_{A,B} = {\arg\max}_{\tau = -\tau_{max} \dots \tau_{max}}  \, \{|r_{A,B}(\tau)|\} \, .
\end{equation}
The corresponding correlation is denoted as $r^*_{A,B}$. 
There are two problems with this approach. First, it is harder to examine the statistical
significance of  $|r^*_{A,B}|$ 
because it is the maximum of a set of random variables.\footnote{An analytic 
approach based on extreme-value statistics was proposed in \cite{kramer2009network} but it relies on several
approximations. Numerical approaches based on frequency-domain bootstrapping, on the other hand, 
are computationally expensive \cite{kramer2009network,martin2014estimating,rummel2010analyzing}.} 
Second, it is often the case that there is a range of lag values that produce ``almost maximum''
cross-correlations, say within one standard deviation from each other. 
Focusing on $\tau^*_{A,B}$ and
ignoring the rest of the statistically significant and almost equal cross-correlations is not well justified.

Instead, we follow a more robust approach in which an edge of the domain-level network $G_{\delta}$ may
be associated with a range of lag values.\footnote{In principle, it may be a set of lag values. In practice
though, significant correlations result for a continuous range of lag values.} 
The lag range that we associate with the edge between $A$ and $B$, denoted as $R_\tau(A,B)$, is
defined as {\em the range of lags that produce significant cross-correlations, within one 
standard deviation from $|r^*_{A,B}|$.} 
If $R_\tau(A,B)$ includes $\tau$=0, the edge is represented as {\em undirected}. 
If $R_\tau(A,B)$ includes only positive lags, the edge is directed from $A$ to $B$
meaning that $A$'s signal precedes $B$'s by the given lag range; otherwise, we associate the 
opposite direction with that edge. 
We emphasize that the directionality of the edges does {\em not} imply causality; it only refers
to temporal ordering.
\\

{\large {\em Edge weight and domain strength.}}
How to assign a weight to each domain-level edge in $G_{\delta}$? A common approach is 
to consider the (signed) magnitude of the cross-correlation $r^*_{A,B}$.
This is reasonable if all domain signals have approximately the same signal power.
In addition, we propose a new  edge weight that is based on the covariance of the two domains:
\begin{equation}
w(A,B) = \mathrm{cov}[X_A(t), X_B(t) ] = \tilde{\sigma}_A \, \tilde{\sigma}_B \, r^*_{A,B} \, . 
\end{equation}
The cross-correlation is computed at lag $\tau^*_{A,B}$ but we could 
use the average of all cross-correlations in $R_\tau(A,B)$ instead.
The weight of an edge can be positive or negative depending on the sign of the
corresponding cross-correlation.

Finally, the strength of a network node (domain) is defined as the sum of the absolute weights of
all edges of that node (ignoring edge directionality).  

\section{Illustration - Comparisons} \label{sec:validation}
The two objectives of this section are to illustrate how the $\delta$-MAPS method works, and 
to contrast the results of the latter with commonly used methods such 
as PCA, ICA, spatial clustering, and overlapping community detection. 
We rely on synthetic data so that the ground-truth is known.

\paragraph*{Synthetic data description}
We construct five domains on a 50$\times$70 spatial grid.
Each domain $i$ is associated with a ``mother'' time series $y_i(t)$, ($i$=1$\dots$5). 
To make the experiment more realistic in terms of autocorrelation structure and 
marginal distribution, each $y_i(t)$ is a real fMRI time series with length $T$=1200 (see \S\ref{sec:appfmri}).
The five mother time series $y_i(t)$ are uncorrelated (absolute cross-correlation $<$0.05 at all lags),
and they are normalized to zero-mean, unit-variance. 
To create correlations between domains (i.e., domain-level edges), 
we construct five new time series $x_i(t)$ based on linear combinations of two or more mother time series. 
For instance, if we set $x_i(t) = (1-\alpha)y_i(t) +\alpha y_j(t+\tau)$ with $0<\alpha<1$ 
and $x_j(t)=y_j(t)$, 
domains $i$ and $j$ become positively correlated at a lag $\tau$; 
the correlation increases with $\alpha$. 
The time series $x_i$ are again normalized to zero-mean, unit-variance. 
We then scale the time series of domain $i$ by a factor $\sqrt{s_i}$ 
to control the variance of each domain ($\mathrm{Var}[x_i(t)]=s_i$). 

For simplicity, each domain is a circle with radius $r_p$. 
A domain has a ``core region'' with the same center and radius $r_c<r_p$; the
core is supposed to be the epicenter of that domain.
Every point in the core has the same signal $x_i(t)$ (before we add random noise). 
Outside the core, the signal attenuates at a distance $d$ from the center of the domain as follows:
\begin{equation} 
x_i(t) = \sqrt{f(d)} \, x_i(t),  \, f(d)=\frac{r_p-d}{r_p-r_c}, \, r_c\leq d \leq r_p \, .
\end{equation} 

\begin{figure*} [ht]
\centering 
\epsfig{file=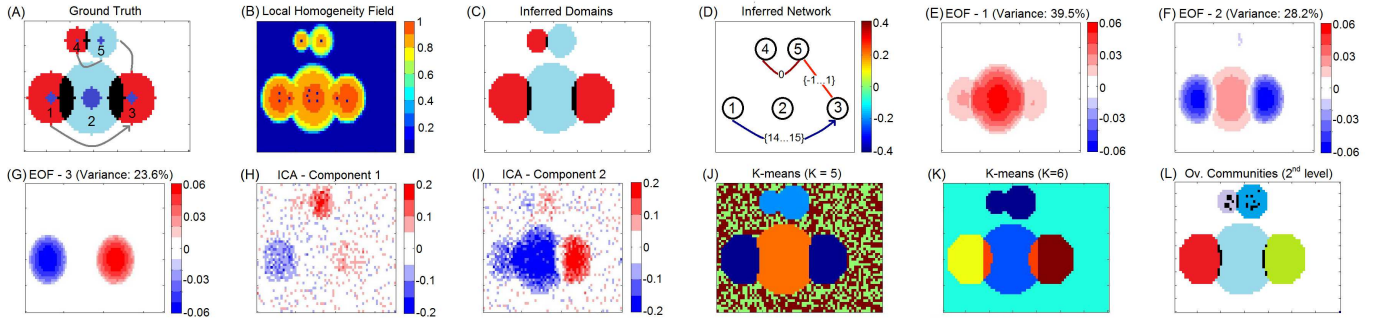,scale=0.18} 
\caption{
{\bf A:} The five ground-truth domains. Adjacent domains have different colors, overlapping regions shown in 
black, and the core of each domain is in blue. The three constructed edges are shown in gray lines. 
{\bf B:} The homogeneity field $\hat{r}_{K}(i)$ at each cell. The identified seeds are shown in blue. 
{\bf C:} The inferred domains: adjacent domains have different colors and overlaps are shown in black.  
{\bf D:} The inferred domain-level network: the color map refers to the edge correlation. The lag associated with each edge is also shown. 
{\bf E,F,G:} The first three EOF (PCA) components. The variance explained by each component is shown at 
the top of each figure. 
{\bf H,I:} The two ICA components. 
{\bf J,K:} K-means clustering.  
{\bf L:} The second hierarchical level of community structure as identified by OSLOM: each community has 
a distinct color and overlaps are shown in black.}
\label{figSynthetic}
\end{figure*}

Finally, we superimpose white Gaussian noise of zero-mean, unit-variance on the entire grid. 
The parameters of the five synthetic domains are shown in Table~\ref{tableSyntheticParams}.
The domains differ in terms of size and power (variance).  
The spatial extent of the domains is shown in Fig.\ref{figSynthetic}-A; 
domains 1 and 3 overlap with domain 2, while domains 4 and 5 also overlap to a smaller extent.
Further, there is a strong and lagged anti-correlation between domains 1 and 3, 
a weaker positive correlation at zero-lag between domains 4 and 5, 
and an ever weaker positive correlation at zero-lag between domains 3 and 5.
The edges of the domain-level network are also shown in Fig.\ref{figSynthetic}-A.

\begin{table}[ht]
\centering
\caption{Synthetic area generation parameters.}
\label{tableSyntheticParams}
\scalebox{1.0}{
\begin{tabular}{|c|c|c|c|c|}
\hline
ID & $r_c$ & $r_p$ & $s_i$ & $x_i(t)$ \\ \hline
1   & 2 & 10 & 16 & $x_1(t)  = 2/3 y_1(t) - 1/3 y_3(t+15)$  \\ \hline
2   & 4 & 14 & 11 & $x_2(t)  = y_2(t)$  \\ \hline
3   & 2 & 10 &16 & $x_3(t)  = y_3(t)$  \\ \hline
4   & 0.5 & 5 & 9 & $x_4(t)  = 3/4 y_4(t) + 1/4 y_5(t)$  \\ \hline
5   & 1    & 7 & 6 & $x_5(t) = 4/5 y_5(t) + 1/5 y_3(t)$  \\ \hline
\end{tabular}
}
\end{table}

\paragraph*{{\large $\delta$}-MAPS results}
The parameters of $\delta$-MAPS are set as follows: $K$=4 cells (up-down-left-right), 
and $\delta$=0.55 (corresponds to significance level $10^{-2}$).
In the edge inference step, the FDR threshold is $q$=10\% and $\tau_{max}=20$.

Fig.\ref{figSynthetic}-B shows the local homogeneity field $\hat{r}_K(i)$ as well as the identified seeds (blue dots), while
Fig.\ref{figSynthetic}-C shows the five discovered domains.
As expected, we often identify more than one seed in the core of each domain due to noise; those
seeds are eventually merged into the same domain. The local homogeneity field is weaker in 
domains 4 and 5 (due to their lower variance) but a seed is still detected in those domains.
Seeds also appear at the two overlapping regions between (1,2) and (2,3) but those seeds gradually 
merge with one of the domains in which they appear.  

Each domain is a subset of the domain's true expanse. 
The reason is that some cells close to the periphery of each domain have very low signal-to-noise ratio (recall
that the signal decays to zero at the periphery and so the average correlation between those cells
with the rest of their domain does not exceed the $\delta$ threshold).    
More quantitatively, the inferred domains include about 80\%-90\% of the ground-truth cells in each domain.
In non-overlapping regions this fraction is higher (85\%-95\% of the cells), while in overlapping
regions it drops to 45\%-80\%. 
The extent of overlapping regions is harder to correctly identify especially when a domain 
(e.g., domain 2) overlaps with a stronger domain (e.g., domains 1 or 3); the stronger domain 
effectively masks the signal of the weaker domain. 
The average pairwise cross-correlation of the cells in each domain varies between 55\%-70\% in 
the ground-truth data, while the inferred domains have slightly higher average cross-correlation
(65\%-75\%) due to their smaller expanse. 

Finally, Fig.~\ref{figSynthetic}-C shows the inferred domain-level network. 
$\delta$-MAPS identifies correctly the three edges and their polarity (positive
versus negative correlations). The lag ranges always include the correct value (e.g., the edge 
between domains 1 and 3 has a lag range [14,15]). Also, the three edges are correctly ordered in terms of 
absolute cross-correlation magnitude: (1,3) followed by (4,5), followed by (3,5).

\paragraph*{PCA/EOF results}
We apply EOF analysis using Matlab's PCA toolbox. 
Fig.~\ref{figSynthetic}-E,F,G show the first three principal components, which collectively 
account for about 90\% of the total variance.
A first observation is that domains 4 and 5 are not even visible in these components -- they only
appear in the next two components, which account for about 5\% of the variance each. This is because
domains 4 and 5 are smaller and have lower variance. This is a general limitation of PCA:
the variance of the analyzed field can be dominated by a small number of ``modes of variability'', 
completely masking smaller/weaker regions of interest and their connections.
Second, the first three components do not provide a consistent evidence that domains 1 and 3 
are strongly anti-correlated; this is due to their lagged correlation, which is missed by PCA.
Third, the first component, which accounts for 40\% of the total variance, can be misinterpreted
to imply that domain 2 is somehow positively correlated with domains 1 and 3, even though it is actually 
generated by an uncorrelated signal. This is due to the overlap of domain 2 with domains 1 and 3.

\paragraph*{ICA results}
We apply ICA on the synthetic data using Matlab's FastICA toolbox. 
To help ICA perform better, we specified the right number of independent components,
which is two (domains 1,3,4,5 are indirectly correlated -- domain 2 is not correlated with any other).
The two independent components are shown in Fig.~\ref{figSynthetic}-H,I. 
Note that only a rough ``shadow'' of each domain is visible. Domains 1 and 3 appear in different colors, 
providing a hint that they are anti-correlated, while domains 3 and 5 appear in the same color
because they are positively correlated. 
Overall, however, the components are quite noisy and it would be hard in practice to discover 
the functional structure of the underlying system if we did not know the ground-truth.   
The results are even harder to interpret when we request a larger number of components.

\paragraph*{Clustering results}
We apply the most well-known clustering method, {\em k-means}, on our synthetic data.
As commonly done with correlation-based clustering, 
the distance between two cells $i$ and $j$ is determined by the maximum absolute correlation across
all considered lags, as $1-|r^*_{i,j}|$.
Fig.~\ref{figSynthetic}-J,K shows the resulting clusters for $k$=5 (the number of 
synthetic domains) and 6, respectively. 
For $k$=5, domains 1 and 3 form a single cluster because of their strong anti-correlation;
the same happens with domains 4 and 5. 
Further, two of the five clusters (green and brown) cover just noise.
The situation changes completely when we request $k$=6 clusters.
In that case, the overlapping regions in domain 2 form a single cluster, while
domains 1 and 3 are separated in different clusters.
Another clustering algorithm, resulting in spatially contiguous clusters \cite{fountalis2014spatio},
is illustrated in \S\ref{sec:appclimate} in the context of climate data analysis (see Fig.~\ref{figEOFSlinksComms}-D).

\paragraph*{Community detection results}
We apply a state-of-the-art overlapping community detection method, referred
to as OSLOM \cite{lancichinetti2011finding}, with the default parameter values.
The input to OSLOM is a positively weighted graph: each vertex is a grid cell and 
an edge between vertices $i$ and $j$ corresponds to the maximum absolute cross-correlation $|r^*_{i,j}|$ 
across all lags of interest. 
Absolute correlations less than 30\% are considered insignificant and the corresponding edges are 
pruned.\footnote{We have experimented with other pruning thresholds between 20\%-50\%  and
the results are very similar at the first two hierarchy levels.}
As most community detection methods, OSLOM 
does not distinguish between positive and negative correlations.
OSLOM provides a hierarchy of communities. 
When applied to our synthetic data, the first level of hierarchy (not shown) simply groups together
domains 1,2,3 in one community (even though domain 2 is uncorrelated with domains 1 and 3), 
and domains 4,5 in another community.  
The connection between domains 3 and 5 is missed.
The second level of hierarchy is shown in Fig.~\ref{figSynthetic}-L.
Overall, OSLOM does a better job than PCA/ICA/clustering in detecting the spatial extent of 
each domain. A small overlap between domains (1,2) and (2,3) is discovered but to 
a smaller extent than $\delta$-MAPS.
However, a community in OSLOM is not constrained to be spatially contiguous. This is the 
reason we see some black dots in regions 4 and 5; these are non-contiguous overlaps between the communities
that correspond to these two domains.

\section{Application in Climate Science} \label{sec:appclimate}
We first apply $\delta$-MAPS in the context of climate science.
Climate scientists are interested in {\em teleconnections} between different regions, 
and they often rely on EOF analysis to uncover them \cite{von2001statistical}.  
Here, we analyze the monthly {\em Sea-Surface Temperature} (SST) field from the HadISST
dataset \cite{rayner2003global}, covering 50 years (1956-2005) at a spatial resolution of $2.0^o \times 2.5^o$,
and we focus on the latitudinal range of $[60^oS;60^oN]$ to avoid sea-ice covered regions.
Following standard practice, we pre-process the time series to form {\em anomalies}, i.e., remove the seasonal cycle, 
remove any long-term trend at each grid-point (using the Theil-Sen estimator),
and transform the signal to zero-mean at each grid point.

$\delta$-MAPS is applied as follows. We set the local neighborhood to the $K$=4 nearest cells so that
we can identify the smallest possible domains at the given spatial resolution.
Second, the homogeneity threshold $\delta$ is set to 0.37 (corresponds to a significance level of $10^{-2}$). 
In the edge inference stage, the lag range is $\tau_{max}$=12 months
(a reasonable value for large-scale changes in atmospheric wave patterns), 
and the FDR threshold is set to $q$=3\% (we identify about 30 edges and so we expect no more than one false positive).

\begin{figure*}[htpb]
\centering
\epsfig{file=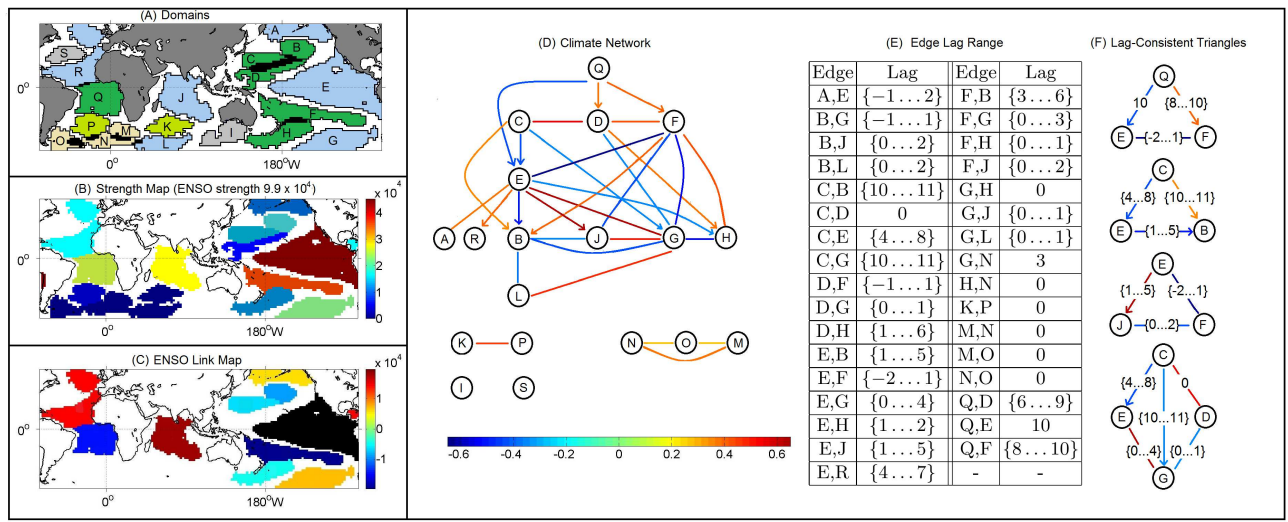,scale=0.13}
\caption{(A) The identified domains. The color of each domain corresponds to the connected
component it belongs to (the blue and green nodes belong to two different poles of the same component).
(B) Color map for domain strength. The strength of ENSO (domain $E$) is shown at the top. 
(C) Edges to and from ENSO (shown in black). 
(D) The climate network. The color of each edge represents the corresponding cross-correlation. 
(E) The lag range associated with each edge.
(F) Examples of lag-constistent triangles.}
\label{figSpatialSimField}
\end{figure*}

Fig.~\ref{figSpatialSimField}-A shows the identified domains (the color code will be explained shortly). 
The spatial dimensionality has been reduced from about 6000 grid cells to 18 domains.
65\% of the sea-covered cells belong to at least one domain; the overlapping regions are shown in black and they 
cover 2\% of the grid cells that belong to a domain.
The largest domain (domain $E$) corresponds to the El Ni$\tilde{n}$o Southern Oscillation (ENSO), 
which is also the most important in terms of node strength (see Fig.~\ref{figSpatialSimField}-B).
Other strong nodes are domain $F$ (part of the ``horseshoe-pattern'' surrounding ENSO), 
domain $J$ (Indian ocean) and domain $Q$ (sub-tropical Atlantic). 
The strength of the edges associated with ENSO are shown 
in Fig.~\ref{figSpatialSimField}-C.
These findings are consistent with known facts in climate science regarding ENSO and its positive correlation 
with the Indian ocean and north tropical Atlantic, and negative correlations with 
the regions that surround it in the Pacific (horseshoe-pattern) \cite{klein1999remote}.  
 
Fig.~\ref{figSpatialSimField}-D shows the inferred domain-level network. 
The color code represents the (signed) cross-correlation for each edge. 
The lag range associated with each edge is shown in Fig.~\ref{figSpatialSimField}-E;
recall that some edges are not directed because their lag range includes $\tau$=0.
The network consists of five weakly-connected components. 
If we analyze the largest component (which includes ENSO) as a signed network (i.e., some edges
are positive and some negative) we see that it is {\em structurally balanced} \cite{easley2010networks}.
A graph is structurally balanced if it does not contain cycles with an odd number
of negative edges.\footnote{For instance, if two friends are both enemies with a third person, 
they form a balanced social triangle.} A structurally balanced network can be partitioned in a ``dipole'', 
so that positive edges only appear within each pole and negative edges appear only between
the two poles. In Fig.~\ref{figSpatialSimField}-A, the nodes of these two poles 
are colored as blue and green (the smaller disconnected components are shown in other colors). 

Focusing on the lag range of each edge, domain $Q$ seems to play a unique role, 
as it temporally precedes all other domains in the inferred network.
Specifically, its activity precedes that of domains $D$, $E$ and $F$ by about 5-10 months. 
The lead of south tropical Atlantic SSTs (domain $Q$) on ENSO has recently received significant attention
in climate science \cite{rodriguez2009atlantic}.
Our results suggest that SST anomalies in domain $Q$ may impact  a large portion of the climate system.


Switching to lag inference, we say that a triangle is {\em lag-consistent} if there is at least one 
value in the lag range associated with each edge 
that would place the three nodes in a consistent temporal distance with respect to each other. 
For instance, in the case of the first triangle of Fig.~\ref{figSpatialSimField}-F, the triangle is 
lag-consistent if the edge from $Q$ to $F$ has a lag of 8 months and the edge between $E$ and $F$ has lag -2 months
(meaning that the direction would be from $F$ to $E$); several other values would make this triangle lag-consistent. 
We have verified the lag-consistency of every triangle in the climate network. 
One exception is the triangle between domains ($C,D,G$), shown at the bottom of Fig.~\ref{figSpatialSimField}-F.
However, the large lag in the edge from $C$ to $G$ can be explained with the triangle
between domains ($C,E,G$), which is lag-consistent.  
We emphasize that the temporal ordering that results from these lag relations should not 
be misinterpreted as causality; we expect that several of the edges we identify are
only due to indirect correlations, not associated with a causal interaction between the corresponding two nodes.

\begin{figure}[htpb]
\centering
\epsfig{file=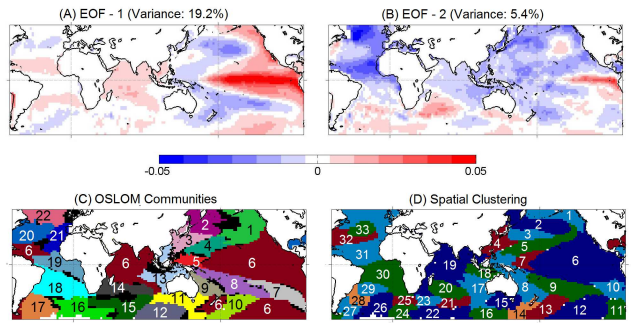,scale = 0.15}
\caption{(A),(B) The first two components of EOF analysis. (C) Communities identified by OSLOM. 
Each community has a unique number and color. (D) Areas identified by spatial clustering. }
\label{figEOFSlinksComms}
\end{figure}

For comparison purposes, Fig.~\ref{figEOFSlinksComms} shows the results of EOF analysis,
community detection, and spatial clustering on the same dataset.
The first EOF explains only about 19\% of the variance, implying that the SST field
is too complex to be understood with only one spatial component. On the other hand, the 
joint interpretation of multiple EOF components is problematic due to their orthogonal relation 
\cite{dommenget2002cautionary}. 
The anti-correlation between ENSO and the horseshoe-pattern regions is well captured in the first 
component but several other important connections, such as the negative and lagged relation between 
the south subtropical Atlantic and ENSO (domains $Q$ and $E$, respectively), are missed. 

Fig.~\ref{figEOFSlinksComms}-C shows the results of the overlapping community detection method OSLOM.  
Following \cite{steinhaeuser2010exploration}, the input to OSLOM is a correlation-based cell-level network. 
Correlations less than $30\%$ are ignored. 
The weight of each edge is set to the maximum absolute correlation  between the corresponding 
two cells, across all considered lags. 
OSLOM identifies $22$ communities.
Community $6$ is not spatially contiguous; it covers ENSO, the Indian ocean, a region 
in the north tropical Atlantic, and a region in south Pacific. 
This is a general problem with community detection methods: 
they cannot distinguish high correlations due to a remote connection 
from correlations due to spatial proximity.
In the context of climate, the former may be due to 
atmospheric waves or large-scale ocean currents while the latter may be due to local circulations.

Finally, Fig.~\ref{figEOFSlinksComms}-D shows the results of a spatial clustering method
\cite{fountalis2014spatio}, with the same homogeneity threshold $\delta$ we use in $\delta$-MAPS. 
That method ensures that every cluster (referred to as ``area'') is spatially contiguous 
but it also requires that there is no overlap between areas and it attempts to assign each grid cell to an area. 
Consequently, it results in more areas (compared to the number of domains),
some of which are just artifacts of the spatial parcellation process.  
Further, the spatial expanse of an area constrains the computation of subsequent areas 
because no overlaps are allowed.

\section{Application in fMRI data} \label{sec:appfmri}
Functional magnetic resonance imaging (fMRI) measures fluctuations of the blood
oxygenation level dependent (BOLD) signal in the brain. 
The dynamics of the BOLD signal in gray matter
are generally correlated with the level of neural activity.
The resulting spatio-temporal field is often analyzed using ICA, clustering or network-based
methods to infer {\em brain functional networks} \cite{sporns2011networks}.

Here, we illustrate $\delta$-MAPS on cortical {\em resting-state} fMRI data from 
a single subject (healthy young male adult, subject-ID: 122620) 
from the WU-Minn Human Connectome Project (HCP) \cite{van2013wu}.  
The data acquisition parameters are described in \cite{smith2013resting}. 
The spatial resolution is 2mm in each voxel dimension.
The pre-processing of fMRI data requires several steps; we use the ``fix-extended''
HCP minimal processing pipeline that includes head motion correction, registration to a structural
image, masking on non-brain voxels, etc; please see \cite{glasser2013minimal}. 
MELODIC ICA and FIX are used to remove non-neuronal artifacts (e.g., physiological noise 
due to cardiac and respiratory cycles). 
We also perform bandpass filtering in the range 0.01-0.08Hz, as commonly done in resting-state fMRI.

In this paper, 
we analyze two scanning runs of the same subject (``scan-1'' and ``scan-2''). 
Each scan lasts about 14 minutes and results in a time series of length $T$=1200 (repetition time TR=720msec).
We emphasize that major differences across different scanning sessions of the same subject
are common in fMRI; studies of functional brain networks often only report group-level averages. 
The entire cortical volume is projected to a surface mesh (Conte69 32K)
resulting in about 65K {\em gray-ordinate} points (as opposed to volumetric voxels) \cite{van2012parcellations}. 
Each point of this mesh is adjacent to six other points; for this reason we set $K$=6.
The homogeneity threshold is set to $\delta$=0.37 (corresponds to significance level $10^{-2}$). 
The maximum lag range $\tau_{max}$ is set to $\pm$3, i.e., 2.2 seconds,
and the FDR threshold is set to $q$=$10^{-4}$ (i.e., we expect one out of 10K edges to be a false positive).
The signal of a domain is defined as the average across all voxels in that domain. 

The application of $\delta$-MAPS results in a network with about 850 domains in scan-1 
(1120 domains in scan-2). 
80\% of the domains are smaller than 30-40 voxels (depending on the scan)
and 5\% of the domains are larger than 250 voxels.  
The number of edges is 4285 in scan-1 (4200 in scan-2).
The absolute value of the cross-correlation associated with each edge is typically larger than 0.5.
The fraction of negative edge correlations is about 5\% in scan-1 and 20\% in scan-2 suggesting
that the polarity of some network edges may be time-varying.
The lag $\tau^*$ that corresponds to the maximum cross-correlation is 0 in 70\% of the edges
and $\pm$1 in almost all other cases. 
13\% of the edges are directed, meaning that lag-0 does not produce a significant correlation for 
that pair of domains. 
There is a positive correlation between the degree of a domain and its physical size
(the correlation coefficient between degree and $\log_{10}$(size) is 0.70 for scan-1
and 0.66 for scan-2).
Further, the network is assortative meaning that domains tend to connect to other domains of
similar degree (assortativity coefficient about 0.7 in both scans).  

\begin{figure*}[htpb]
\centering
\epsfig{file=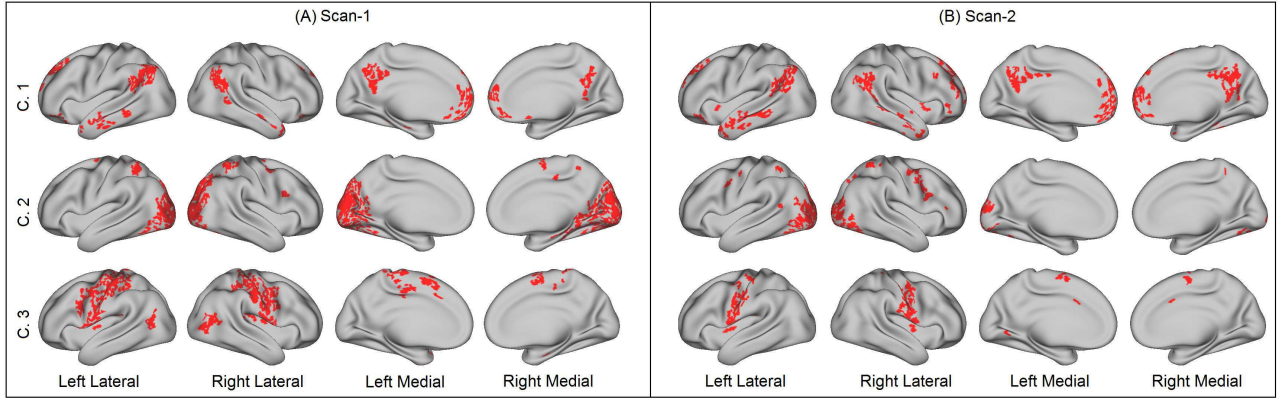,scale = 0.14}
\caption{Three domain-level network communities for each scan. The first corresponds to 
the default-mode network, the second to the occipital network, and the third to 
the motor/somatosensory network. }
\label{figFMRIcomms}
\end{figure*}
An important question is whether the $\delta$-MAPS networks are consistent with what neuroscientists currently
know about resting-state activity in the brain. 
During rest, certain cortical regions that are collectively referred to as the {\em Default-Mode Network (or DMN)}
are persistently active across age and gender \cite{yeo2011organization}. 
Other known resting-state networks are the occipital (part of the visual system)
and the motor/somatosensory (associated with planning and execution of voluntary body motion). 
With the terminology of network theory, the previous ``networks'' would be referred to as 
{\em communities} within the larger functional brain network. 
To identify communities in the $\delta$-MAPS network, we applied OSLOM \cite{lancichinetti2011finding}.
OSLOM identifies two hierarchical levels in both scans. 
The first level consists of highly overlapping communities that cover almost the entire cortex.
The second hierarchical level is more interesting, resulting in eight communities for scan-1 (nine  for scan-2). 
Fig.~\ref{figFMRIcomms} shows the three communities (C.1, C.2, C.3) for each scan that have the highest 
resemblance to the three previously mentioned resting-state networks: 
C.1 corresponds to the DMN,
C.2 corresponds to the occipital resting-state network, 
and C.3 corresponds to the motor/somatosensory network.
C.1 is quite similar across the two scanning sessions and it clearly captures the DMN.
In C.2, the  extent of the network is smaller in scan-2, which is not too surprising giving the 
known inter-scan variability of resting-state fMRI.
C.3 is also quite similar across the two scans and consistent with the motor/somatosensory network.

\begin{figure}[htpb]
\centering
\epsfig{file=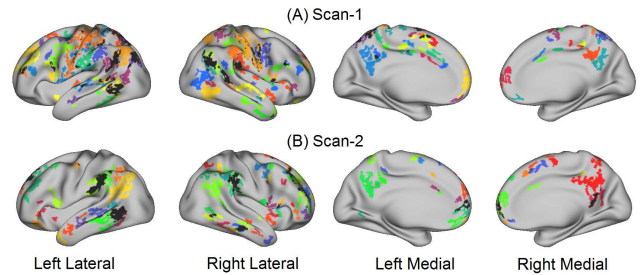,scale = 0.14}
\caption{The domains of the backbone network for each hemisphere and scan. The color of each domain is randomly assigned (overlaps are shown in black).} 
\label{fig:richclub}
\end{figure}

To further investigate the structure of those higher degree (and typically larger) domains, we perform
{\em k-core decomposition}.\footnote{A process that starts with the original network ($k$=0), 
and it removes iteratively all nodes of degree $k$ or less in each round 
so that after the extraction of the $k$'th core all remaining nodes have degree larger than $k$.}
The density of the remaining network, after the extraction of $k$=14 cores from the scan-1 network
($k$=16 cores in scan-2) shows a sudden increase by a factor of two.
This suggests that the network includes a {\em densely inter-connected backbone}, also known as ``rich-club''.
The size of this backbone is small relative to the entire network: 130 domains in scan-1 (90 in scan-2).
Similar observations about the resting-state brain, but using voxel-level network analysis methods,
have been previously reported \cite{van2011rich}. 
Fig.\ref{fig:richclub} shows the location of the backbone domains for each hemisphere and for each scan.
The regions that are usually associated with the DMN dominate the backbone of both sessions. 
Interestingly though, scan-1 includes the regions of the motor/somatosensory network, 
while the backbone of scan-2 is missing those regions. 
One possible explanation for this discrepancy is that the subject was more relaxed during 
scan-2, not exerting the mental effort to stay still. 

\section{Discussion} \label{sec:conclusions}
$\delta$-MAPS results in a correlation-based functional network. 
A next step would be to infer a causal, or {\em effective} network, leveraging the framework of
probabilistic graphical models. Instead of attempting to learn the graph structure from raw data, 
one could use the $\delta$-MAPS network as the underlying structure and then apply 
conditional independence tests to remove non-causal edges (e.g., \cite{ebert2014causal}).
Additionally, in many real systems the underlying temporal dynamics are non-stationary. 
Instead of relying on sliding window-based approaches, which are often sensitive to the 
duration of the window,  an important extension of $\delta$-MAPS will be to construct 
dynamic networks by detecting automatically the time periods during which the network remains constant. 
It would also be interesting to combine the inferred functional network with a structural
network that shows the physical connectivity between the identified domains. 
This is not hard in the case of communication networks but it becomes also feasible
for brain networks using diffusion-weighted MRI. 
The projection of the observed dynamics on the underlying structure can help to characterize 
the actual function and delay of each system component. 





\tiny{

}

\section*{Appendix I: Identifying the \\ largest domain is NP-complete} \label{appendixI}

{\normalsize

We are given a spatio-temporal field $\mathbf{X}(t)$ on a grid $G$, a pairwise similarity metric between pairs of grid cells and a threshold $\delta$. Starting from a grid cell $c$, the goal is to find the largest subset of grid cells that form a single spatially connected component, and whose average similarity exceeds the threshold $\delta$. The spatial grid can be represented as a planar graph $G(V,E)$ where each grid cell is a node and edges connect adjacent grid cells. Formally we have the following graph optimization problem: \newline

{\emph{Definition 1.} Rooted Largest Connected $\delta$-Dense Subgraph Problem (rooted LC$\delta$DS). Given a regular (grid) graph $G(V,E)$, a weight function $w: V\times V \rightarrow \mathbb{R}$ (where $w(v,v)=0$ and symmetric), a threshold $\delta$, and a node $c \in V$, find a maximum cardinality set of nodes $A \subseteq V $ such that $c \in A$, the induced subgraph is connected ($I_G(A) = 1$) and $\frac{\sum_{v,u\in A}w(v,u)}{|A|(|A|-1)} > \delta$ (i.e., $\hat{r}(A) > \delta$). \newline

To show that rooted LC$\delta$DS is NP-hard we first consider a variant of the problem in which the induced subgraph $A$ has to satisfy two conditions; it has to be a connected subgraph of $G$, and the average weight of the edges in $A$ has to exceed $\delta$. More formally: \newline

{\emph Definition 2.} Largest Connected $\delta$-Dense Subgraph Problem (LC$\delta$DS).  Given a regular (grid) graph $G(V,E)$, a weight function $w: V\times V \rightarrow \mathbb{R}$ (where $w(v,v)=0$ and symmetric), and a threshold $\delta$,  find a maximum cardinality set of nodes $A\subseteq V$ such that $I_G(A) = 1$ and $\hat{r}(A) > \delta$. \newline

To show that LC$\delta$DS is NP-hard we use a reduction of the densest connected $k$ subgraph problem. \newline

{\emph Definition 3.} Densest Connected $k$-Subgraph Problem \\ (DC$k$S). { Decision version: Given a graph $G(V,E)$,  and positive integers $k$ and $j$, does there exist an induced subgraph on $k$ vertices such that this subgraph has at least $j$ edges and is connected? \newline

DCkS (also referred to as the connected h-clustering problem) has been shown to be NP-complete on general graphs \cite{corneil1984clustering}, as well as on planar graphs \cite{keil1991complexity}. DC$k$S is polynomially time solvable for subclasses of planar graphs of bounded tree width \cite{arnborg1991easy}. 
Grid graphs, which are the type of graphs that arise in our application domains, are planar bipartite graphs, with non-fixed tree width, and no positive results are known for this subclass of planar graphs. 
The work on approximating densest/heaviest connected k-subgraphs is relatively very limited (see recent theoretical result \cite{chen2015finding}).  It is easy to show that the DC$k$S problem can be easily reduced to an instance of the decision version of the LC$\delta$DS problem, and hence it is also NP-complete even on planar graphs. \newline

LEMMA 1. The decision version of the LC$\delta$DS problem is NP-complete on planar graphs. \newline

PROOF. This can be shown via a reduction from the DC$k$S.  We reduce an instance $<G,k,j>$ of the DC$k$S to an LC$\delta$DS instance by using the same graph $G$, setting $w(u,v)=I(u,v)\in E$ ($w(u,v)$ is 1 if and only if the pair of nodes is connected by an edge), and $\delta=j/k(k-1)$. \newline

Now it is easy to show that rooted LC$\delta$DS is also NP-hard. If a poly-time algorithm existed for the rooted LC$\delta$DS, then by calling it $|V|$ times with each of the nodes of the graph, we would obtain in poly-time a solution to the NP-hard LC$\delta$DS.
}

\section*{Appendix II: Heuristic for the selection of {\large{$\delta$}}} 
{\normalsize
The threshold $\delta$ intuitively determines the minimum degree of homogeneity that the underlying field must have within each domain.
The higher the threshold, the higher the required homogeneity and therefore, the smaller the size of the identified domains.

To select $\delta$ we propose the following heuristic. We start with a random sample of pairs of grid cells and for each pair  $i,j$ we compute the Pearson correlation $r_{i,j}$ at zero lag. To assess the significance of each correlation we use Bartlett's formula \cite{box2011time}. Under the null hypothesis of no coupling $r_{i,j}$ should have zero mean, and a reasonable estimate of its variance is given by 

\begin{equation}\label{EqBarlettVarianceAreas}
Var[ r_{i,j} ] = \frac{1}{T} \sum_{\tau_k = -T}^T r_{i,i} (\tau_k) r_{j,j}(\tau_k)~,
\end{equation}

here $r_{i,i}(\tau_k)$ is the autocorrelation of the time series of grid cell $i$ at lag $\tau_k$. The scaled values $z_{i,j} = \frac{ r_{i,j} } { \sqrt{ Var[r_{i,j} ] } }$ should approximately follow a standard normal distribution. To assess the significance of each correlation we perform a one sided z-test for a given level of significance $\alpha$. 

The threshold $\delta$ is set as the average of all significant correlations. A domain is a set of spatially contiguous grid cells, thus we require that the mean pairwise correlation for the cells belonging to the same domain to be higher than the mean pair-wise correlation of randomly picked pairs of grid cells. $\delta$ depends on the choice of the significance level $\alpha$, on the autocorrelation structure of the underlying time series and on the correlation distribution of the field.
 
}

\section*{Appendix III: {\large{$\delta$}}-MAPS pseudocode} \label{appendixIII}

\makeatletter
\renewcommand{\Function}[2]{%
  \csname ALG@cmd@\ALG@L @Function\endcsname{#1}{#2}%
  \def\jayden@currentfunction{#1}%
}
\newcommand{\funclabel}[1]{%
  \@bsphack
  \protected@write\@auxout{}{%
    \string\newlabel{#1}{{\jayden@currentfunction}{\thepage}}%
  }%
  \@esphack
}
\makeatother

\begin{algorithm}[!htpb]
\begin{algorithmic}[1]

\State Domains $S=\{A_1, \dots ,A_{|S|}\}$ \Comment{The initial set of domains}
\Function{DomainIdentification}{\null}
\While{True}
 \State boolean $merged \gets $ \Call{DomainMerging}{$S$}
 \State boolean $expanded \gets $ \Call{DomainExpansion}{$S$}
 \If{$!merged \&\& !expanded$}
	\State break \Comment{Terminate when no further expansion or merging is possible}
 \EndIf
\EndWhile	
	
\EndFunction

\end{algorithmic}
\end{algorithm}

\begin{algorithm}[!htpb]
\begin{algorithmic}[1]

\Function{DomainExpansion}{Domains $S=\{A_1, \dots ,A_{|S|}\}$} \funclabel{alg1}
	\State boolean $startMerging \gets false$
	\State boolean $expanded \gets false$
	\While{$!startMerging$} \Comment{Domain expansion is repeated in rounds}
		\State $expanded \gets false$
		\State sort($S$) \Comment{Sort domains in decreasing order of homogeneity such that $\hat{r}(A_{i-1}) > \hat{r}(A_i) > \hat{r}(A_{i+1})$ }
		\For{$i = 1:|S|$}
			\State Domain $A_i \gets S[i]$ 
			\State Domain $eA_i \gets $ \Call{ExpandDomain}{$A_i$} \
			\If{$|A_i| \neq  |eA_i|$} \Comment{Domain  expanded}
				\State $S[i] \gets eA_i$
				\State $expanded \gets true$
				\State $startMerging \gets $ \Call{CanMerge}{$eA_i$} 
				\If{$startMerging$}
					\State break \Comment{Exit the for loop}
				\EndIf
			\EndIf
		\EndFor 
		\Comment{A round of domain expansion is complete}
		\If{$!expanded$} 
			\State break \Comment{Domains cannot be expanded}
		\EndIf
	\EndWhile
	\State \Return $expanded$
\EndFunction

\State

\Function{ExpandDomain}{Domain $A_i$}
		\Comment{Try do expand domain $A_i$ by one cell}
		\State Construct set $\Gamma(A_i)$: all cells adjacent to $A_i$
		
		\If{$\Gamma(A_i) = \emptyset$}
			\State \Return $A_i$
		\Else
			\State $m \gets \arg\max_{m \in \Gamma(A_i)} \hat{r}(A_i \cup \{m\})$	\Comment{Select the cell that maximizes $\hat{r}(A_i \cup \{m\})$.}				
			\If{$\hat{r}(A_i \cup \{m\}) > \delta$}
				\State $A_i  \gets  A_i \cup m$
			\EndIf
			\State \Return $A_i$
		\EndIf		
\EndFunction

\State

\Function{CanMerge}{Domain $A_i$}
\Comment{Check whether one or more merging operations are possible}
\State boolean $merge \gets false$
\State Construct set $\Gamma(A_i)$: all domains adjacent to  $A_i$
\For{$j = 1:| \Gamma(A_i) |}$
	\State $A_j \gets \Gamma(A_i)[j]$
	\If{$ \hat{r}(A_i \cup A_j) > \delta $}
		\State $merge \gets true$
		\State break
	\EndIf
\EndFor
\State \Return $merge$
\EndFunction

\end{algorithmic}
\end{algorithm}

\begin{algorithm}[!htpb]
\begin{algorithmic}[1]

\Function{DomainMerging}{Domains $S=\{A_1, \dots ,A_{|S|}\}$}
\State boolean $merged \gets false$
\While{True} \Comment{Repeat until no pair of domains can be merged}
	\State Domain $DomainToMerge1 \gets \emptyset$
	\State Domain $DomainToMerge2 \gets \emptyset$ \Comment{Domains with the maximum union homogeneity}
	\State $maxHomogeneity \gets -1$ 
	\For{$i = 1:|S|$} 
		\State Domain $A_i \gets S[i]$ \Comment{Get the $i^{th}$ domain}
		\State Construct set $\Gamma(A_i)$
		\State $A_j \gets \arg\max_{A_j \in \Gamma(A_i)} \hat{r}(A_i \cup A_j)$
		

		\If{$\hat{r}(A_i \cup A_j) > maxHomogeneity$ }  \Comment{Update the best candidates to merge}
			\State $DomainToMerge1 \gets A_i$
			\State $DomainToMerge2 \gets A_j$
			\State $maxHomogeneity \gets \hat{r}(A_i \cup A_j)$ 
		\EndIf		
	\EndFor
	\If{$maxHomogeneity > \delta$}
			\State S.remove($DomainToMerge1$) 
			\State S.remove($DomainToMerge2$) \Comment{Remove the domains that will be merged}
			\State S $\gets DomainToMerge1 \cup DomainToMerge2$ 
			\State $merged \gets true$

	\Else
		\State{break} \Comment{We can not merge any domains}
	\EndIf
\EndWhile	
\State \Return $merged$ \Comment{Return true if at least one pair of domains is merged}

\EndFunction

\end{algorithmic}
\end{algorithm}


\begin{thebibliography}{1}


\bibitem{ahn2010link}
Y.~Y. Ahn, J.~P. Bagrow and  S.~Lehmann, ``Link communities reveal multiscale complexity in networks,'' \hskip 1em plus
  0.5em minus 0.4em\relax {\emph Nature}, 466(7307):761–764, 2010.

\bibitem{arnborg1991easy} S.~Arnborg, J.~Lagergren, and D.~Seese, ``Easy problems for
tree-decomposable graphs,''\hskip 1em plus
  0.5em minus 0.4em\relax {\emph Journal of Algorithms},
12(2):308-340, 1991.

\bibitem{baldassano2015parcellating}
C.~Baldassano, D.~M. Beck, and L.~Fei-Fei, ``Parcellating connectivity in spatial maps,'' \hskip 1em plus
  0.5em minus 0.4em\relax {\emph PeerJ}, 3:e784, 2015.

\bibitem{benjamini1995controlling}
Y.~Benjamini and Y.~Hochberg, ``Controlling the false discovery rate: {A} practical and powerful approach to multiple testing,'' \hskip 1em plus
  0.5em minus 0.4em\relax {\emph J. R. Stat. Soc.
Series B}, pages 289–300, 1995.


\bibitem{blumensath2012resting}
T.~Blumensath, T.~E. Behrens, and S.~M. Smith, ``Resting-state FMRI single subject cortical parcellation based on region growing,'' \hskip 1em plus
  0.5em minus 0.4em\relax {\emph MICCAI 2012}, pages
188–195. Springer, 2012.


\bibitem{blumensath2013spatially}
T.~Blumensath, S.~Jbabdi, M.~F. Glasser, D.~C. Van Essen,
K.~Ugurbil, T.~E. Behrens, and S.~M. Smith, ``Spatially
constrained hierarchical parcellation of the brain with
resting-state fMRI,'' \hskip 1em plus
  0.5em minus 0.4em\relax {\emph Neuroimage}, 76:313–324, 2013.

\bibitem{box2011time}
G.~E. Box, G.~M. Jenkins, and G.~C. Reinsel, ``Time series
analysis: forecasting and control,'' \hskip 1em plus
  0.5em minus 0.4em\relax volume 734. John Wiley
\& Sons, 2011.


\bibitem{chen2015finding} X.~Chen, X.~Hu, and C.~Wang, ``Finding connected dense
k-subgraphs,'' \hskip 1em plus
  0.5em minus 0.4em\relax {\emph In Theory and Applications of Models of
Computation}, pages 248-259. Springer, 2015.

\bibitem{corneil1984clustering} D.~G. Corneil and Y.~Perl, ``Clustering and domination in
perfect graphs,'' \hskip 1em plus
  0.5em minus 0.4em\relax {\emph Discrete Applied Mathematics}, 9(1):27-39,
1984.

\bibitem{craddock2012whole}
R.~C. Craddock, G.~A. James, P.~E. Holtzheimer, X.~P. Hu,
and H.~S. Mayberg, ``A whole brain fMRI atlas generated via
spatially constrained spectral clustering,'' \hskip 1em plus
  0.5em minus 0.4em\relax {\emph Hum. Brain Mapp.}, 33(8):1914–1928, 2012.

\bibitem{dommenget2002cautionary}
D.~Dommenget and M.~Latif, ``A cautionary note on the
interpretation of EOFs,'' \hskip 1em plus
  0.5em minus 0.4em\relax {\emph J. of Climate}, 15(2):216–225,
2002.


\bibitem{donges2009backbone}
J.~F. Donges, Y.~Zou, N.~Marwan, and J.~Kurths, ``The
backbone of the climate network,'' \hskip 1em plus
  0.5em minus 0.4em\relax {\emph EPL}, 87(4):48007, 2009.

\bibitem{easley2010networks}
D.~Easley and J.~Kleinberg, ``Networks, crowds, and
markets: Reasoning about a highly connected world,'' \hskip 1em plus
  0.5em minus 0.4em\relax Cambridge University Press, 2010.

\bibitem{ebert2014causal}
I.~Ebert-Uphoff and Y.~Deng, ``Causal discovery from
spatio-temporal data with applications to climate science,'' \hskip 1em plus
  0.5em minus 0.4em\relax {\emph ICMLA 2014}, pages 606–613. IEEE,
2014.


\bibitem{fountalis2014spatio}
I.~Fountalis, A.~Bracco, and C.~Dovrolis, ``Spatio-temporal
network analysis for studying climate patterns,'' \hskip 1em plus
  0.5em minus 0.4em\relax {\emph Climate
Dynam.}, 42(3-4):879–899, 2014.

\bibitem{glasser2013minimal}
M.~F. Glasser, S.~N. Sotiropoulos, J.~A. Wilson, T.~S.
Coalson, B.~Fischl, J.~L. Andersson, J.~Xu, S.~Jbabdi,
M.~Webster, J.~R. Polimeni, et al., ``The minimal
preprocessing pipelines for the Human Connectome
Project,'' \hskip 1em plus
  0.5em minus 0.4em\relax {\emph Neuroimage}, 80:105–124, 2013.


\bibitem{hyvarinen1999fast}
A.~Hyv{\"a}rinen, ``Fast and robust fixed-point algorithms for independent component analysis,'' \hskip 1em plus
  0.5em minus 0.4em\relax  {\emph IEEE Trans. Neural Netw.}, 10(3):626–634, 1999.

\bibitem{keil1991complexity} J.~M. Keil and T.~B. Brecht, ``The complexity of clustering
in planar graphs,'' \hskip 1em plus
  0.5em minus 0.4em\relax {\emph J. Combinatorial Mathematics and
Combinatorial Computing}, 9:155-159, 1991.

\bibitem{kawale2013graph}
J.~Kawale, S.~Liess, A.~Kumar, M.~Steinbach, P.~Snyder, V.~ Kumar,
A.~R. Ganguly, N.~F. Samatova, F.~Semazzi, ``A graph-based
approach to find teleconnections in climate data,'' \hskip 1em plus
  0.5em minus 0.4em\relax {\emph Stat Anal Data
Min}, 6(3):158-179, 2013.

\bibitem{klein1999remote}
S.~A. Klein, B.~J. Soden, and N.~-C. Lau, ``Remote sea
surface temperature variations during ENSO: Evidence for
a tropical atmospheric bridge,'' \hskip 1em plus
  0.5em minus 0.4em\relax {\emph J. Climate},
12(4):917–932, 1999.

\bibitem{kramer2009network}
M.~A. Kramer, U.~T. Eden, S.~S. Cash, and E.~D. Kolaczyk, ``Network inference with confidence from multivariate time
series,'' \hskip 1em plus
  0.5em minus 0.4em\relax {\emph Phys. Rev. E}, 79(6):061916, 2009.

\bibitem{lancichinetti2011finding}
A.~Lancichinetti, F.`Radicchi, J.~J. Ramasco, abd S.~Fortunato, ``Finding statistically significant communities in
networks,'' \hskip 1em plus
  0.5em minus 0.4em\relax  {\emph PloS One}, 6(4):e18961, 2011.

\bibitem{lu2003region}
Y.~Lu, T.~Jiang, and Y.~Zang, ``Region growing method for
the analysis of functional MRI data,'' \hskip 1em plus
  0.5em minus 0.4em\relax   {\emph NeuroImage}, 20(1):455–465, 2003.

\bibitem{martin2014estimating}
E.~Martin and J.~Davidsen, ``Estimating time delays for
constructing dynamical networks,'' \hskip 1em plus
  0.5em minus 0.4em\relax  {\emph Nonlinear Proc. Geoph.}, 21(5):929–937, 2014.

\bibitem{mcguire2014community}
M.~P. McGuire and N.~P. Nguyen, ``Community structure
analysis in big climate data,'' \hskip 1em plus
  0.5em minus 0.4em\relax  in {\emph IEEE International Conference on Big Data, 2014}, pages 38–46. IEEE,
2014.

\bibitem{palla2005uncovering}
G. Palla, I. Der{\'e}nyi, I. Farkas, and T. Vicsek, ``Uncovering
the overlapping community structure of complex networks in nature and society,'' \hskip 1em plus
  0.5em minus 0.4em\relax {\emph Nature}, 435(7043):814–818, 2005.

\bibitem{power2011functional}
J.~D. Power, A.~L. Cohen, S.~M. Nelson, G.~S. Wig, K.~A.
Barnes, J.~A. Church, A.~C. Vogel, T.~O. Laumann, F.~M.
Miezin, B.~L. Schlaggar, et al., ``Functional network
organization of the human brain,'' \hskip 1em plus
  0.5em minus 0.4em\relax  {\emph Neuron}, 72(4):665–678,
2011.

\bibitem{rayner2003global}
N.~Rayner, D.~E. Parker, E.~Horton, C.~Folland,
L.~Alexander, D.~Rowell, E.~Kent, and A.~Kaplan, ``Global
analyses of sea surface temperature, sea ice, and night
marine air temperature since the late nineteenth century,'' \hskip 1em plus
  0.5em minus 0.4em\relax {\emph J. Geophys. Res.: Atmospheres
(1984–2012)}, 108(D14), 2003.

\bibitem{reiner}
A.~Reiner, D.~Yekutieli, and Y.~Benjamini, ``Identifying differentially expressed genes using false discovery rate controlling procedures,'' \hskip 1em plus 0.5em minus 0.4em\relax {\emph Bioinformatics} 19: 368-375, 2003.


\bibitem{rodriguez2009atlantic}
B.~Rodr{\'\i}guez-Fonseca, I.~Polo, J.~Garc{\'\i}a-Serrano,
T.~Losada, E.~Mohino, C. R.~Mechoso, and F.~Kucharski, ``Are {A}tlantic {N}i{\~n}os enhancing {P}acific {ENSO} events in recent decades?,'' \hskip 1em plus
  0.5em minus 0.4em\relax {\emph Geophys. Res. Lett.}, 36(20), 2009.

\bibitem{rummel2010analyzing}
C.~Rummel, M.~M{\"u}ller, G.~Baier, F.~Amor, and
K.~Schindler, ``Analyzing spatio-temporal patterns of
genuine cross-correlations,'' \hskip 1em plus
  0.5em minus 0.4em\relax  {\emph J. Neurosci. 
methods}, 191(1):94–100, 2010.

\bibitem{smith2013resting}
S.~M. Smith, C.~F. Beckmann, J.~Andersson, E.~J.
Auerbach, J.~Bijsterbosch, G.~Douaud, E.~Duff, D.~A.
Feinberg, L.~Griffanti, M.~P. Harms, et al., ``Resting-state
fMRI in the human connectome project,'' \hskip 1em plus
  0.5em minus 0.4em\relax  {\emph Neuroimage},
80:144–168, 2013.

\bibitem{sporns2011networks}
O.~Sporns, ``Networks of the Brain,'' \hskip 1em plus
  0.5em minus 0.4em\relax  MIT press, 2011.

\bibitem{steinbach2003discovery}
M.~Steinbach, P.~-N. Tan, V.~Kumar, S.~Klooster, and
C.~Potter, ``Discovery of climate indices using clustering,'' \hskip 1em plus
  0.5em minus 0.4em\relax  in {\emph ACM SIGKDD}, 2003, pages
446–455.

\bibitem{steinhaeuser2010exploration}
K.~Steinhaeuser, N.~V. Chawla, and A.~R. Ganguly, ``An
exploration of climate data using complex networks,'' \hskip 1em plus
  0.5em minus 0.4em\relax  {\emph ACM
SIGKDD Explorations Newsletter}, 12(1):25–32, 2010.

\bibitem{steinhaeuser2011complex}
K.~Steinhaeuser, N.~V. Chawla, and A.~R. Ganguly, ``Complex networks as a unified framework for descriptive
analysis and predictive modeling in climate science,'' \hskip 1em plus
  0.5em minus 0.4em\relax  {\emph Stat. Anal. Data Min.}, 4(5):497–511, 2011.

\bibitem{thirion2014fmri}
B.~Thirion, G.~Varoquaux, E.~Dohmatob, and J~.-B. Poline, ``Which fMRI clustering gives good brain parcellations?,'' \hskip 1em plus
  0.5em minus 0.4em\relax  {\emph Data Front. Neurosci.}, 8, 2014.

\bibitem{van2008normalized}
M.~Van Den Heuvel, R.~Mandl, and H.~Hulshoff Pol, ``Normalized cut group clustering of resting-state fMRI data,'' \hskip 1em plus
  0.5em minus 0.4em\relax  {\emph PloS One}, 3(4):e2001, 2008.


\bibitem{van2011rich}
M.~P. van den Heuvel and O.~Sporns, ``Rich-club
organization of the human connectome,'' \hskip 1em plus
  0.5em minus 0.4em\relax {\emph J. Neurosci.}, 31(44):15775–15786, 2011.

\bibitem{van2012parcellations}
D.~C. Van Essen, M.~F. Glasser, D.~L. Dierker, J.~Harwell,
and T.~Coalso, ``Parcellations and hemispheric asymmetries
of human cerebral cortex analyzed on surface-based atlases,'' \hskip 1em plus
  0.5em minus 0.4em\relax  {\emph Cereb. Cortex}, 22(10):2241–2262, 2012.

\bibitem{van2013wu}
D.~C. Van Essen, S.~M. Smith, D.~M. Barch, T.~E. Behrens,
E.~Yacoub, K.~Ugurbil, W.~-M. H. Consortium, et al, ``The
WU-Minn human connectome project:An overview,'' \hskip 1em plus
  0.5em minus 0.4em\relax  {\emph Neuroimage}, 80:62–79, 2013.

\bibitem{von2001statistical}
H.~Von Storch and F.~W. Zwiers, ``Statistical analysis in
climate research,'' \hskip 1em plus
  0.5em minus 0.4em\relax  Cambridge university press, 2001.


\bibitem{yeo2011organization}
B.~T. Yeo, F.~M. Krienen, J.~Sepulcre, M.~R. Sabuncu,
D.~Lashkari, M.~Hollinshead, J.~L. Roffman, J.~W. Smoller,
L.~Z{\"o}llei, J.~R. Polimeni, et al, ``The organization of the
human cerebral cortex estimated by intrinsic functional
connectivity,'' \hskip 1em plus
  0.5em minus 0.4em\relax {\emph J. Neurophysiol.}, 106(3):1125–1165,
2011.



\end{thebibliography}
\end{document}